# Defect-Fluorite $Gd_2Zr_2O_7$ Ceramics under Helium Irradiation: Amorphization, Cell Volume Expansion, and Multi-stage Bubble Formation

Zhangyi Huang [1,2,3,4], Nannan Ma [1], Jianqi Qi [1,2,5,*], Xiaofeng Guo [3,6], Mao Yang [1,2], Zhe Tang [1], Yutong Zhang[1], Yong Han[7], Di Wu [3,4,6,8,*], Tiecheng Lu[1,2,5,*],

[1] College of Physical Science and Technology, Sichuan University, Chengdu 610064, China

[2] Key Laboratory of Radiation Physics and Technology of Ministry of Education, Sichuan University, Chengdu 610064, China

[3] Alexandra Navrotsky Institute for Experimental Thermodynamics, Washington State University, Pullman, Washington 99164, United States

[4] The Gene and Linda Voiland School of Chemical Engineering and Bioengineering, Washington State University, Pullman, Washington 99164, United States

[5] Key Laboratory of High Energy Density Physics of Ministry of Education, Sichuan University, Chengdu 610064, China

[6] Department of Chemistry, Washington State University, Pullman, Washington 99164, United States

[7] Department of Physics and Astronomy, Iowa State University, Ames, Iowa 50011, United States

[8] Materials Science and Engineering, Washington State University, Pullman, Washington 99164, United States

---

[*]Corresponding authors email address: qijianqi@scu.edu.cn (J. Qi), d.wu@wsu.edu (D. Wu) and lutiecheng@scu.edu.com (T. Lu).

**Abstract**

Here, we report a study on the radiation resistance enhancement of $Gd_2Zr_2O_7$ nanograin ceramics, in which amorphization, cell volume expansion and multi-stage helium (He) bubble formation are investigated and discussed. $Gd_2Zr_2O_7$ ceramics with a series of grain sizes (55 – 221 nm) were synthesized and irradiated by 190 keV He ion beam up to a fluence of $5x10^{17}$ ions/cm$^2$. Both the degree of post irradiation cell volume expansion and the amorphization fraction appear to be size dependent. As the average grain size evolves from 55 to 221 nm, the degree of post irradiation cell volume expansion increases from 0.56 to 1.02 %, and the amorphization fraction increases from 6.8 to 11.1 %. Additionally, the threshold He concentrations (at.%) of bubbles at different formation stages and locations, including (1) bubbles at grain boundary, (2) bubble-chains and (3) ribbon-like bubbles within the grain, are all found to be much higher in the nanograin ceramic (55 nm) compared with that of the submicron sample (221 nm). We conclude that grain boundary plays a critical role in minimizing the structural defects, and inhibiting the multi-stage He bubble formation process.

**Introduction**

Crystalline $A_2B_2O_7$ oxides with defect fluorite or pyrochlore phases have great potential to be applied as synthetic rocks for high-level nuclear waste immobilization (SYNROC). Radioactive actinides may be confined and immobilized include Pu, Am, and Cm [1-3]. However, typically, under highly radioactive environments, the irradiation-induced amorphization damage the ceramic waste form structures resulting in poor chemical durability and/or radioactive nuclide release.

The irradiation resistance properties of $A_2B_2O_7$ materials to amorphization have been systematically studied [4-6]. $Gd_2Zr_2O_7$ is often considered as a radiation tolerant $A_2B_2O_7$ oxide, which cannot be completely amorphized even at dose up to 100 dpa [3-5]. At a fluence of $1\times10^{16}$ $Xe^{20+}$/cm, more than 25 % $Gd_2Zr_2O_7$ is amorphized [2,7]. In addition, other than the deleterious effect induced by structural damage, He atoms undergo a continuous accumulation due to α-decay, forming He bubbles in the $Gd_2Zr_2O_7$ host matrix, and causing swelling and surface blistering [8,9]. Moreover, He accumulation may also generate mechanical stress, which results in cracking, thereby increasing the interfacial contact of radionuclides with ground water, and reducing the chemical durability of the waste forms [8].

Recently, nanostructured materials with high grain boundary density were found to be promising waste form candidates to minimize the detrimental radiation defects induced by accumulation of structural damage and impurity (He atoms). For instance, enhanced radiation resistance to amorphization was observed for nanostructured $MgGa_2O_4$, yttria stabilized zirconia (YSZ), and $Gd_2(Ti_{0.65}Zr_{0.35})_2O_7$ powder[10-14]. Additionally, improved radiation tolerance to swelling and hardening caused by He bubble was also observed for nanostructured steel, tungsten film, and Fe–Cr–Ni alloy [15-17]. Nevertheless, thus far, the exact role of grain boundary and/or grain size in (1) structural orderness or amorphization, (2) cell volume expansion and (3) bubble formation induced by He irradiation for the $A_2B_2O_7$ ceramic, remains unclear.

Here, we present a study on He ion radiation resistance behavior of $Gd_2Zr_2O_7$ ceramics to amorphization, unit cell swelling and multi-stage He bubble formation as the sample grain size increases from nanoscale (55 nm) to submicron scale (221 nm). The starting material we used, nanocrystalline $Gd_2Zr_2O_7$ powder, has a pure defect fluorite phase. It was synthesized by a solvothermal assisted co-precipitation method followed by calcination at 800 °C in air for 4 h [18]. Subsequently, the as obtained $Gd_2Zr_2O_7$ powders were consolidated into ceramic pellets by field assisted sintering technique (FAST) using a LABOX-325 system (Sinter Land, Japan). We determined the average grain sizes of samples sintered at 1170 °C (55 nm), 1240 °C (70 nm), 1320 °C (123 nm) and 1400 °C (221 nm) for 5 min using a statistical measurement from fractured cross-section images employing Nano Measure software. These $Gd_2Zr_2O_7$ samples with a series of grain sizes were double-side polished using fine metallographic abrasive paper and diamond paste until reaching surface roughness less than 10 nm.

**Experimental Methods**

$Gd_2Zr_2O_7$ ceramics were irradiated with 190 keV He ion beam to the fluence of $5 \times 10^{17}$ ions/cm$^2$ at room temperature using a 200 keV ion implanter (LC22-100-01, Beijing Zhongkexin Electronics Equipment Co., Ltd). Peak damage (displacement per atom) and He concentration (atomic percentage) were examined by SRIM 2008 (Stopping and Range of Ions in Matter, 2008) using threshold displacement energies of 72 (Gd), 121 (Zr) and 41 (O) eV [19]. Grazing-incidence X-ray diffraction (GIXRD) measurement (Bruker AXS D8 Advances X-ray diffractometer with Cu Ka radiation, λ=1.5406 Å) was conducted at an incident angle of 6° according to Taylor's results to investigate the phase evolution of the samples upon irradiation [8]. The non-irradiated side of each ceramic pellet was also examined with GIXRD under the same experimental condition. Whole-pattern profile fitting of the diffraction data was performed to obtain the lattice parameters, diffraction peak areas, and FWHM (full width at half maximum) of each sample before and after irradiation. Focused ion beam (FIB) was employed to prepare the cross-sections of the irradiated samples. Transmission electron microscopy (TEM, FEI Tecnai $G^2$ F20 TWIN) was performed to elucidate the post He irradiation sample microstructural evolutions.

**Results and Discussions**

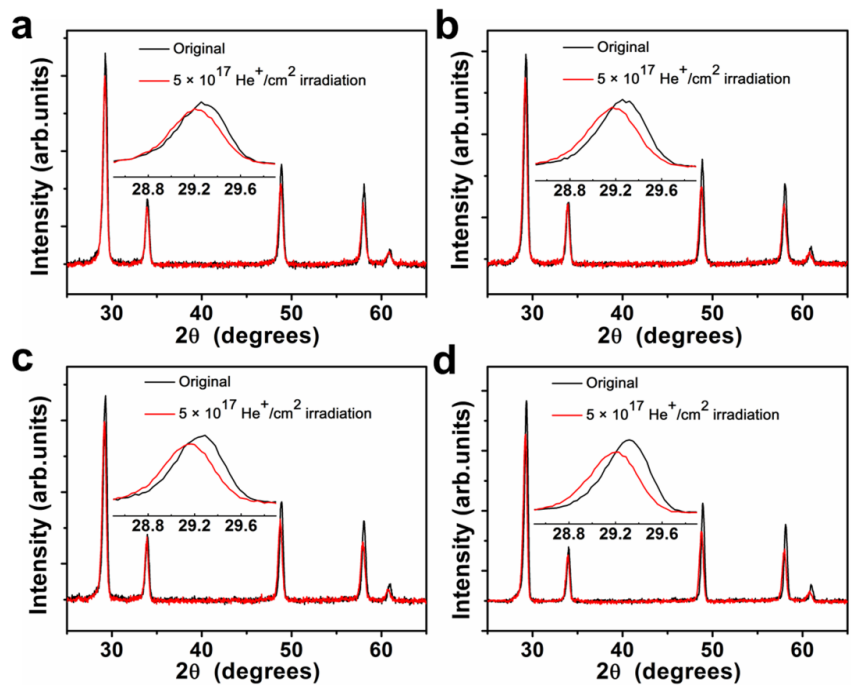

**Figure. 1.** GIXRD patterns of $Gd_2Zr_2O_7$ ceramics prior and post He irradiation: (a) 55, (b) 70, (c) 123, and (d) 221 nm.

Figure 1 shows the GIXRD patterns of $Gd_2Zr_2O_7$ ceramics (55 – 221 nm) before and after He irradiation at $5 \times 10^{17}$ ions/cm$^2$. All the samples share a pure defect-fluorite phase before irradiation, and such structure is well-retained after irradiation. Interestingly, left shift was observed for all irradiated ceramic pellets, indicating lattice expansion induced by He implantation. In addition, intensity reduction and peak broadening, owing to irradiation-induced amorphization [11] were observed for all post irradiation samples.

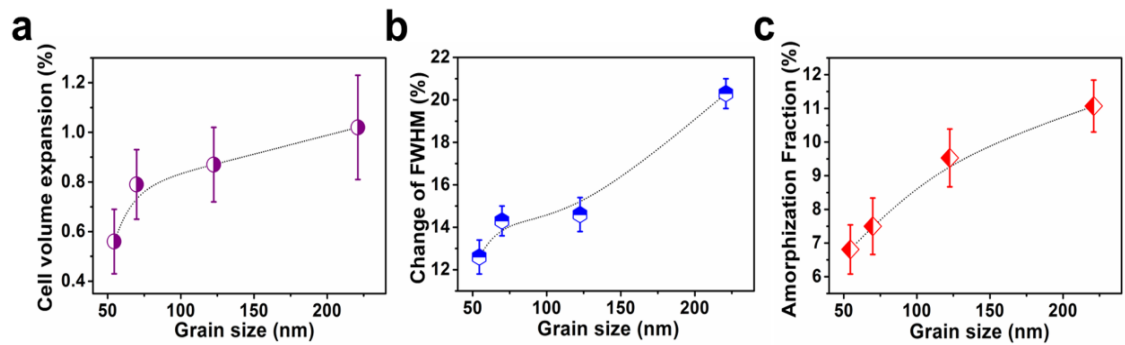

**Figure. 2.** Post irradiation (a) cell volume expansion, (b) FWHM increase, and (c) degree of amorphization for $Gd_2Zr_2O_7$ ceramics as grain size varies.

Interestingly, left shifts, intensity reduction and peak broadening appear to be grain size-dependent (see Figure 2). We calculated the lattice parameter increase and cell volume expansion. Surprisingly, as the average grain size increases from 55 to 221 nm, the degree of post irradiation cell volume expansion increases from 0.56 to 1.02 % (see Figure 2a). Additionally, the FWHM increases 12.6 % for nanograin (55 nm) ceramic, and 20.3 % for the sub-micron ceramic (221 nm, see Figure 2b). It is very likely that the diffuse scattering induced by formation of amorphous phase during irradiation process account for the peak broadening [20]. Further, we derived the amorphous phase fraction ($f_A$) from the net area of diffraction peak using the following equation [7]:

$$f_A = 1 - \frac{\sum_{i=1}^{n} \frac{A_i^{irradiated}}{A_i^{unirradiated}}}{n}$$

$A_i^{irradiated}$ and $A_i^{unirradiated}$ represent the areas of the $i^{th}$ GIXRD peaks of the irradiated and non-irradiated sides, respectively. $n$ is the number of diffraction peaks. The post-irradiation amorphous phase fraction is plotted as a function of average grain size in Figure 2c. Specifically, for the 55 and 221 nm samples, 6.8 and 11.1 % of the crystalline phase were amorphized after irradiation, respectively.

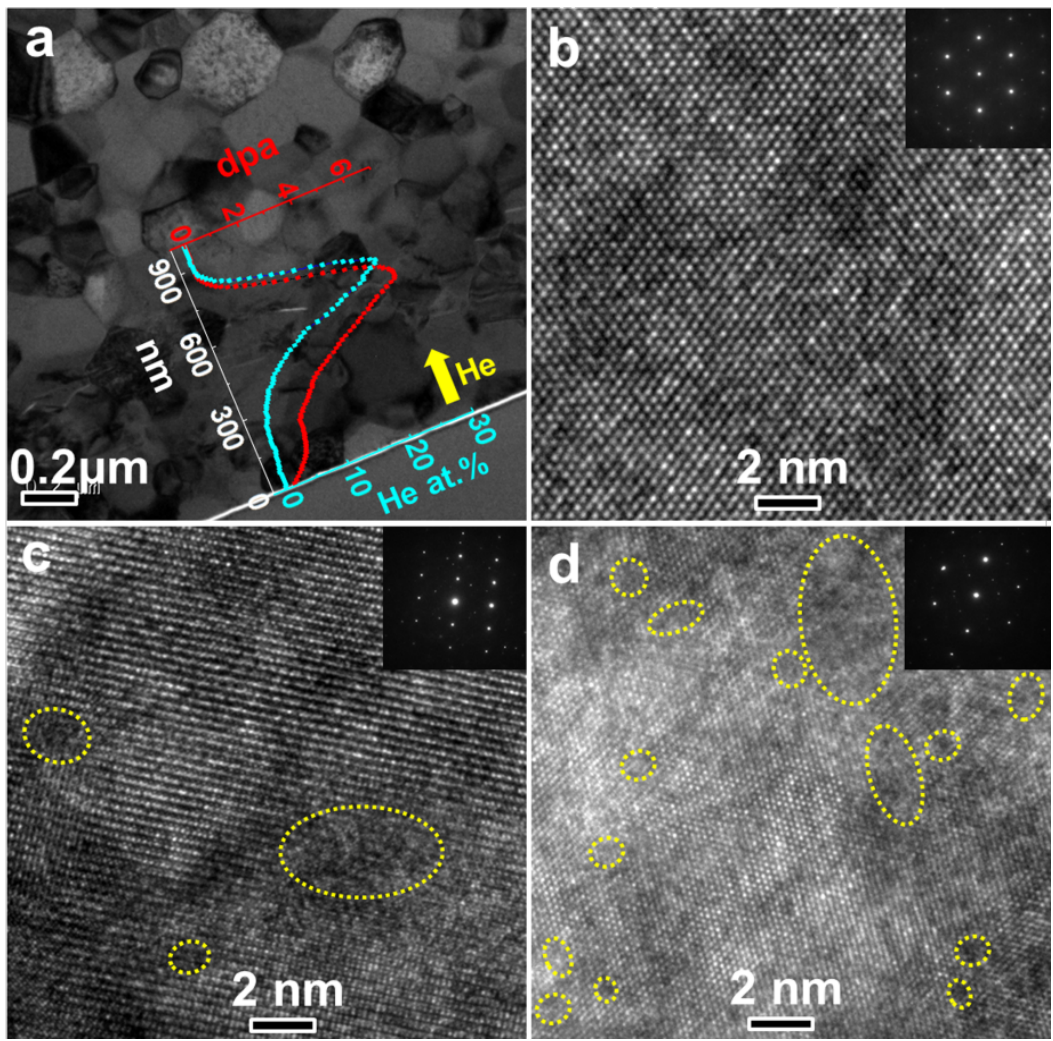

**Figure. 3.** TEM images of the post-He irradiation sub-micron $Gd_2Zr_2O_7$ ceramics: (a) cross-sectional microstructure, and HRTEM images and corresponding SAED patterns recorded from (b) the non-irradiated sample, (c) the upper irradiated layer (~1.8 dpa) and (d) the peak damaged layer (~6.3 dpa). The inset of Figure 3a shows the depth profiles of radiation damage in dpa and the He concentration (at.%) obtained from SRIM simulation.

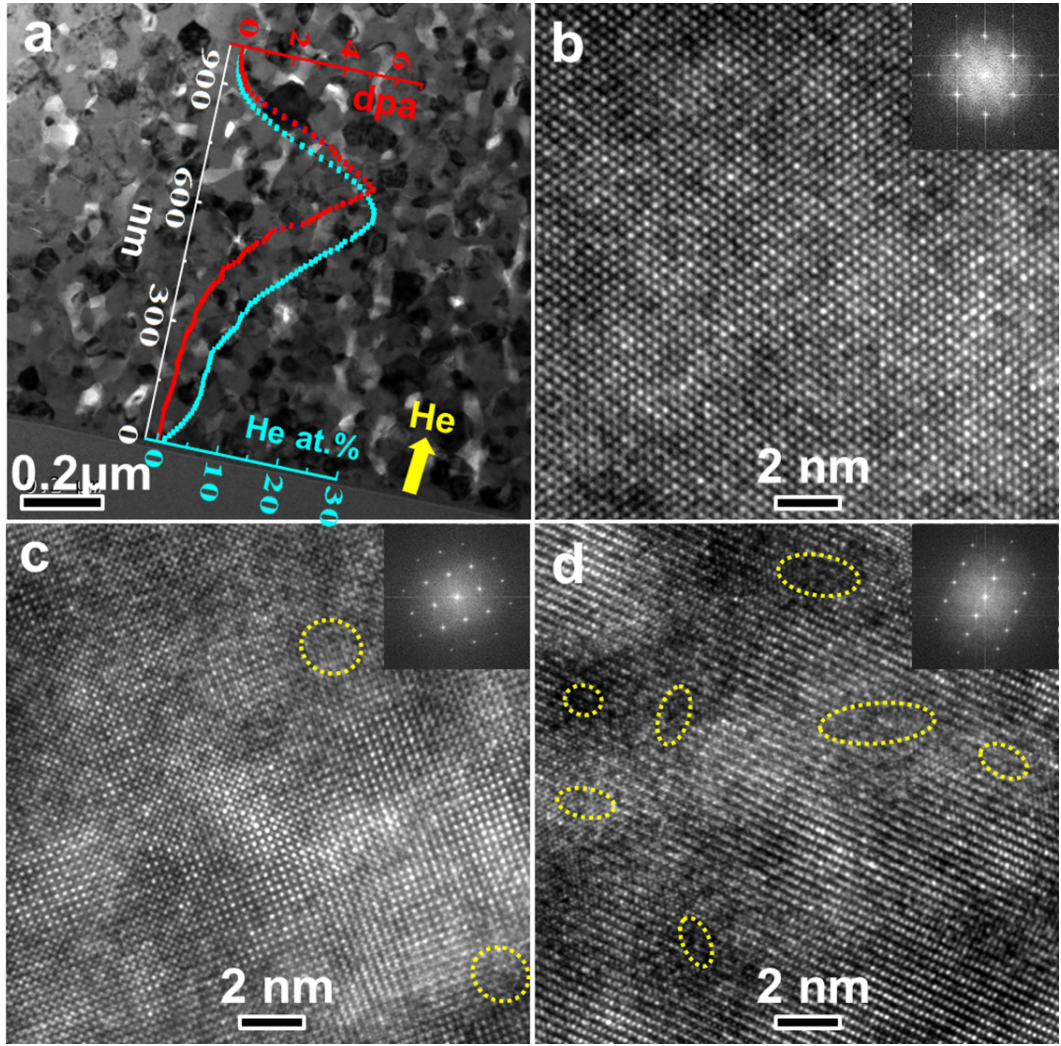

**Figure. 4.** TEM images of the post-He irradiation nanograin $Gd_2Zr_2O_7$ ceramics: (a) cross-sectional microstructure, and HRTEM images and corresponding SAED patterns recorded from (b) the non-irradiated sample, (c) the upper irradiated layer (~1.8 dpa) and (d) the peak damaged layer (~6.3 dpa). The inset of Figure 3a shows the depth profiles of radiation damage in dpa and the He concentration (at.%) obtained from SRIM simulation.

TEM was employed to investigate the microstructure size dependence and He bubble formation of $Gd_2Zr_2O_7$ ceramics upon He irradiation (see Figure 3 and 4). For the sub-micron ceramic sample, a clearly damaged region can be observed (see Figure 3a). Interestingly, in the TEM image of the nanograin ceramic sample, such damage region is not as clearly resolved as that of the sub-micron sample (see Figure 4a). The SAED patterns and FFT images presented in Figure 3b to d and Figure 4b to d suggest that both the sub-micron and nanograin $Gd_2Zr_2O_7$ retain their crystalline structures (partial order structure) at a dose up to 6.4 dpa. However, in the HRTEM images of non-irradiated substrate, the upper irradiated layer (about 1.8 dpa) and the peak damaged layer (about 6.3 dpa) of both the submicron and the nanograin ceramics, the radiation defects mainly present in the form of vacancy. The concentration of vacancy-type defects or defect clusters increases as the dpa value increases. Meanwhile, the average size of defect cluster (see the yellow circles in Figure 3 and 4) clearly increases in the submicron sample yet appears to be the same in nanograin ceramic. Moreover, the increasing local concentration (number) and size of vacancy clusters can be distinguished as grain size increases. According to an earlier study, the vacancy-type defects dominate the major radiation defect forms inside grains is because of the much weaker mobility compared with that of interstitials, which are eventually absorbed by the grain boundaries or are migrated into the material surface [21-23]. Furthermore, the total defect concentration increases as the grain size increases, in which the grain boundaries act as neutralizer for both vacancies and interstitials [22].

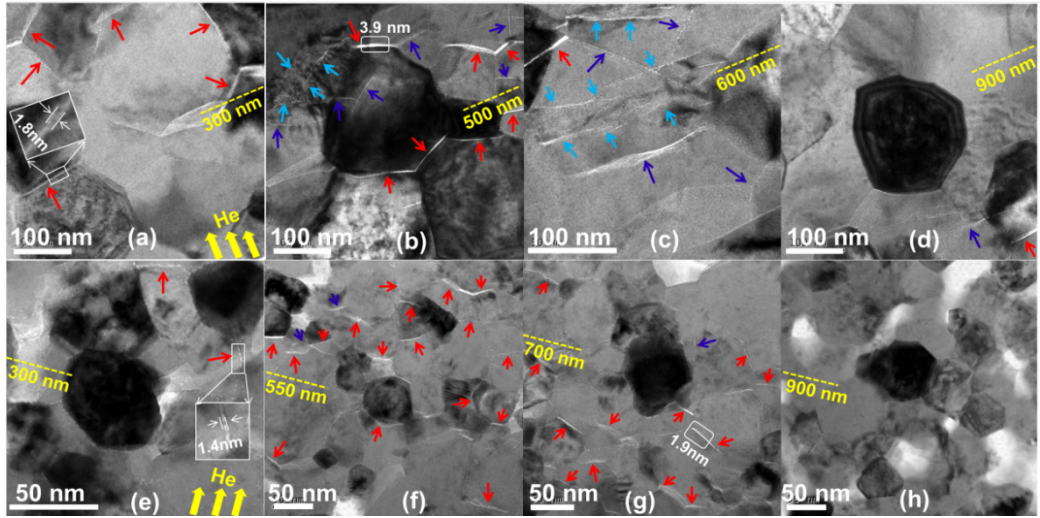

**Figure 5.** TEM images recorded at different depths (away from the surface) for post-He irradiation $Gd_2Zr_2O_7$ ceramics: (a), (b), (c) and (d) 221 nm; (e), (f), (g) and (h) 55 nm. The yellow dash lines are parallel to the external sample surface. The red, dark blue and cyan arrows point to (1) He bubble along grain boundary, (2) He bubble chain inside grain and (3) ribbon-like He bubble inside grain, respectively.

The detail behaviors of implanted He in submicron and nanograin ceramics were presented in Figure 5. In a thermodynamic sense, grain boundary is a preferred location to trap implanted He atoms [10]. Specifically, grain boundary segregation is first observed at ~300 nm away from the surface, which increases the width of grain boundary to ~ 1.8 nm (corresponding to 1.7 at.% He in the SRIM calculated concentration profile, see Figure 5a and inset). Such distinct difference originates from coalescence behavior of He bubbles which leads to the formation of interconnected channels parallel with the grain boundary [24] (see the red arrows in Figure 5). Within the near-surface region (~300 nm), the inner grains and grain boundaries which are perpendicular to the external surface are bubble-free. Only the inner grain appears to have no He bubble at approximately 300 nm away from external particle surface. Around the peak damaged area (see Figure 5b and c), the formation of He bubble chains (see dark blue arrows) and ribbon-like He bubbles inside grains (see cyan arrows) was observed at depths of ~435 and ~590 nm (corresponds to 5.9 and 19.2 at.% He), respectively. Besides, the maximum width of the grain boundary with continuously distributed He bubbles increases to 3.9 nm, and the width of the He bubble chains inside grains is about 1.4 nm (measured by Nano Measure software). The unirradiated substrate appears to have no bubbles or bubble chains (see Figure 5d). Further, the behavior of He bubbles in nanograin and submicron ceramics have four major distinct differences (highlighted in Figure 5e to h): (i) Nanograin sample has much higher He concentration leading to formation of He bubble along grain boundary, He bubble chain inside grain, and ribbon-like He bubble inside grain, respectively. We also present corresponding TEM images and calculated SRIM profile, and corresponding results obtained from TEM images and SRIM calculated profile are listed in Table 1. (ii) Nanograin ceramic features much more He bubble along the grain boundary around peak damaged area; (iii) Nanograin sample posseses much thinner maximum grain boundary width distributed with He bubbles (~1.9 nm). (iv) There is no ribbon-like He bubble within nanograin ceramic.

Table 1 Threshold He concentrations (at.%) in submicron and nanograin $Gd_2Zr_2O_7$ ceramics to form multi-stage He bubble.

| Average grain size | 221 nm | 55 nm |
|---|---|---|
| He bubble along grain boundary | 1.7 % | 2.8 % |
| He bubble chain inside grain | 5.9 % | 15.2 % |
| Ribbon-like He bubble | 19.2 % | >25.8 % |

Shen *et al.* [25] proposed a thermodynamic model, in which the differences in radiation damage behavior for nanograin and coarse-grained materials were interpreted and discussed. It is concluded that radiation damage is governed by the simultaneous and integrated effects of healing kinetics and thermodynamic stability [25]. On the one hand, the free energy of system increases as the grain size decreases. Such free energy depression decrease the energetic difference between the crystalline and corresponding amorphous phases, and significantly increases the probability of radiation induced phase transitions (amorphization). One example is that nanocrystalline $ZrO_2$ (< 10 nm) is found to have stronger driving force towards amorphization [26]. On the other hand, grain boundary can act as sink for point defects, promoting efficient localized annihilation of radiation damage via interstitial emission [21]. Once the interfacial area of grain boundary is not significantly enough to impact the material stability, self-healing effect may counter-balance the instability induced by the grain boundary energy. As reported earlier, nanocrystalline $MgGa_2O_4$ retains its crystallinity upon irradiation at 96 dpa. In contrast, bulk $MgGa_2O_4$ surfers from amorphization even at 24 dpa [10]. Considering such integrative effects, it is very likely that there is a grain size on nanoscale which leads to the best radiation resistance performance for $Gd_2Zr_2O_7$ nanoceramics. According to the GIXRD and TEM results, 55 nm appears to be the optimized size for the radiation resistance of $Gd_2Zr_2O_7$ to amorphization.

Other than grain size dependence, nanocrystalline materials also exhibit radiation-induced grain growth phenomena, which typically trigger material life-time decreases in radioactive environments [27]. For instance, the grain size of YSZ ceramic increases from 38 to 45 nm after treatment under irradiation, suggested by well-resolved GIXRD peak sharpening and direct TEM evidences [11]. Moreover, inter-granular cracks were observed in severely damaged region when nanograin samples (25 or 38nm) were irradiated [11]. However, such phenomena were not observed in our study even under high fluence ($5 \times 10^{17}$ ions/cm$^2$), supported by the GIXRD peak broadening and directly identified micron-morphology in TEM images.

Comparing with inner grains, grain boundaries are favored nucleation sites for He bubbles. Such irradiation induced bubbles prefer alignment and distribution along the grain boundaries [17]. As the local concentration of He increases, the He bubble formation mechanism appears to have three distinct stages, including (1) He bubble formation along grain boundary, (2) He bubble chain formation inside grain and (3) ribbon-like He bubble formation inside grain. This phenomenon is in excellent agreement with what was seen for YSZ ceramic [24]. Furthermore, the threshold He concentrations (at.%) for each bubble formation stage is found to be much higher for nanograin $Gd_2Zr_2O_7$ ceramics compared with the submicron samples. In other words, the nanograin ceramics have stronger capability to inhibit the formation and evolution of He bubbles. We conclude that one reason for hysteretic He bubble formation in nanograin ceramics is that the higher grain boundary area enables effective He bubble distribution. On the other hand, the inner grain bubbles are generated by trapping He in structural voids [28]. Owing to the enhanced vacancy defect neutralization by larger grain boundary areas, the void density is effectively reduced, leading to less small size He bubbles.

In our ongoing and future work, we plan to carefully examine the energetic stability (formation enthalpies) of fresh and post-irradiation $Gd_2Zr_2O_7$ ceramics studied in this work using various calorimetric techniques.

**Conclusions**

In summary, the He irradiation behavior of $Gd_2Zr_2O_7$ ceramics with grain size from 55 to 221 nm was studied. Our results suggest that the irradiation induced amorphization and cell volume expansion behaviors of $Gd_2Zr_2O_7$ ceramics are grain size dependent. Specifically, the nanograin ceramic (55 nm) presents the least degree of amorphization and cell volume expansion upon irradiation. Additionally, nanograin ceramics exhibit enhanced capability to inhibit multi-stage He bubble formation. As the average grain size evolves from 55 to 221 nm, the threshold He concentrations (at.%) for (1) bubble formation along grain boundary, (2) bubble chain inside grain and (3) ribbon-like bubble inside grain increase from 1.7 to 5.9 % and 19.2 to 2.8 %, 15.2 to more than 25.8 %, respectively. Compared with the submicron samples, much more He bubbles were observed to be enriched along the grain boundaries, and much thinner maximum width of the grain boundary were found to have He bubbles distributed for the in the nanograin ceramic. Moreover, irradiation induced grain growth and inter-granular cracks were not observed for $Gd_2Zr_2O_7$ nanograin ceramics. We identified that 55 nm is the optimized size for $Gd_2Zr_2O_7$ nanograin ceramics with enhanced radiation tolerance and thermodynamic stability. Our study provides needed fundamental experimental data and general guidance to design and synthesize radiation-resistant ceramic materials applied in extremely radioactive environments.

**Acknowledgements**

This work was supported by the National Natural Science Foundation of the People's Republic of China under Grant NOs. 11505122 and 11775152, the Science and Technology Innovation Team of Sichuan province under Grant NO.15CXTD0025, and the Key Applied Basic Research (2017JY0329) of Sichuan province. D.W. acknowledges the institutional funds from the Gene and Linda Voiland School of Chemical Engineering and Bioengineering at Washington State University.